\documentclass[11pt,twoside,letterpaper]{article} 
\usepackage{times,fancyhdr}
\usepackage[dvips]{graphicx}

\usepackage{pause}
\usepackage{mpmulti}
\usepackage{amsmath,amssymb,graphicx,epsfig,cite,fancyhdr,amsthm,bm}
\usepackage{tikz}
\usepackage{multicol}
\usetikzlibrary{calc}
\usepackage{pdfsync}
\usepackage{multirow}
\usetikzlibrary{calc}

\usepackage{amsfonts}
\usepackage{amsthm,amscd}
\usepackage{amsbsy}
\usepackage{latexsym}
\usepackage{url} 			
\usepackage{layout}
\usepackage{pslatex}
\usepackage{fleqn} 		
\usepackage{makeidx}
\makeindex 						
\usepackage{layout} 	
\usepackage{epstopdf}	
\usepackage{color}

\usepackage[T1]{fontenc}
\usepackage{poligraf} 
\usepackage[letter,cam,center]{crop} 
\usepackage{type1cm} 	
\usepackage{courier}	
\usepackage{lscape} 	


\makeatletter
\def\cleardoublepage{\clearpage\if@twoside \ifodd\c@page\else%
    \hbox{}%
    \thispagestyle{empty}%
    \newpage%
    \if@twocolumn\hbox{}\newpage\fi\fi\fi}

\newcommand*\dashline{\rotatebox[origin=c]{90}{$\dabar@\dabar@\dabar@\dabar@\dabar@\dabar@\dabar@\dabar@$}}

\def\figurename{Figure}
\makeatletter
\renewcommand{\fnum@figure}[1]{\figurename~\thefigure.}
\makeatother

\def\tablename{Table}
\makeatletter
\renewcommand{\fnum@table}[1]{\tablename~\thetable.}
\makeatother

\newtheorem{lem}{Lemma}

\theoremstyle{remark}
\newtheorem{rem}{Remark}

\theoremstyle{definition}

\begin{document}
\title{
{\begin{flushleft}
\vskip 0.45in
{\normalsize\bfseries\textit{Chapter~8}}
\end{flushleft}
\vskip 0.45in
\bfseries\scshape Cognitive Radio Networks: An Information Theoretic Perspective}}
\author{\bfseries\itshape Mojtaba Vaezi\thanks{E-mail address: mvaezi@princeton.edu}\\
Princeton University}
\date{}
\maketitle
\thispagestyle{empty}
\setcounter{page}{1}
\thispagestyle{fancy}
\fancyhead{}
\fancyhead[L]{In: Book Title \\
Editor: Editor Name, pp. {\thepage-\pageref{lastpage-01}}} 
\fancyhead[R]{ISBN 0000000000  \\
\copyright~2007 Nova Science Publishers, Inc.}
\fancyfoot{}
\renewcommand{\headrulewidth}{0pt}

\vspace{2in}



\pagestyle{fancy}
\fancyhead{}
\fancyhead[EC]{Authors}
\fancyhead[EL,OR]{\thepage}
\fancyhead[OC]{Article Name}
\fancyfoot{}
\renewcommand\headrulewidth{0.5pt}

\label{lastpage-01}

\begin{abstract}

 Information-theoretic limits of {\it cognitive radio} networks have been under exploration since more than a decade ago.
Although such limits are  unknown for many networks,  including the simplest case with
two pairs of transmitter-receiver, there are several cases for which the capacity limits are obtained either exactly or up to a  constant gap.
The goal of this chapter is to provide  insights into the nature of  transmission techniques associated with optimal
communication when cognitive radio technology is used. 
Outlining the  state of the art in the information-theoretic analysis of different cognitive systems, we
highlight the salient features/points of the capacity-achieving or capacity-approaching strategies
that should be considered in wireless network design paradigms based on this  technology.
In particular, we  emphasize  on the interaction of cognitive radio with  emerging technologies for 5G networks.
\end{abstract}


\section{Introduction}\label{sec:intro}

 \textit{Cognitive radios} are intelligent  communication devices that  exploit  information about their environment
 to increase the \textit{spectral efficiency} of communication over a given spectrum band.
 Cognitive radio communication is one of the promising technologies for  improving spectrum utilization
 in the fifth generation (5G) of wireless communication systems.
With an eye toward 5G networks, this chapter surveys the fundamental limits of communication and associated transmission
 techniques for various wireless network design paradigms based on this promising technology.

The idea of cognitive radios was born out of the
spectrum shortage in the form of various solutions in which new devices were
allowed to exploit the spectrum of coexisting noncognitive devices while impacting noncognitive users'
communication only minimally.
Cognitive radios  sense their environment, employ advanced radio and signal processing
techniques and use novel spectrum allocation policies to  improve
spectral utilization by concurrently transmitting  or interweaving their signals with
those of existing users.

\begin{figure}
  \centering
  \includegraphics [scale=0.75] {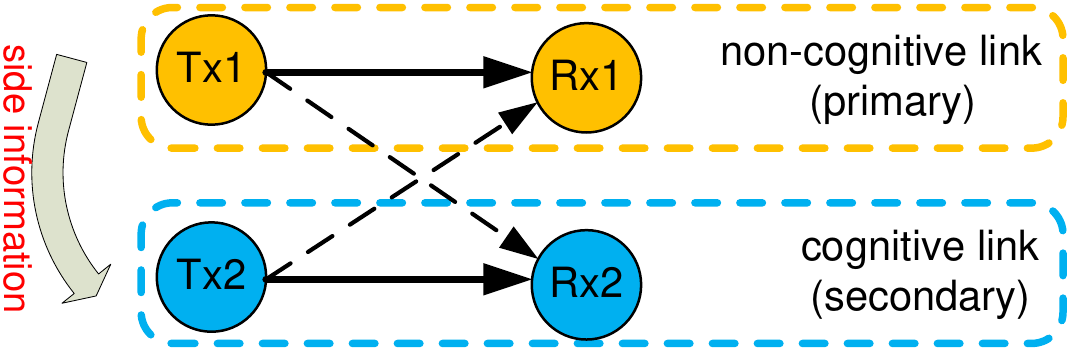}
  \caption{A cognitive network with two pairs of transmitter-receiver. }
  \label{fig:cognitive}
\end{figure}

From an information-theoretic perspective, ``awareness'' of a cognitive node
about other nodes is abstracted as  \textit{side information} which can be
any information about those nodes activity (transmission/reception time), channels state information (CSI), messages, codebooks, etc.
Cognitive communication is then referred to a communication system in which
each cognitive node can make use of any \textit{side information} about
other nodes with which it has a shared spectrum. Figure~\ref{fig:cognitive}
models the simplest cognitive radio network in which there is
one noncognitive transmitter (Tx1)  and one cognitive transmitter (Tx2)
as well as their corresponding receivers (Rx1 and Rx2).
Note that, in Figure~\ref{fig:cognitive}, the direction of side information
is what differentiates the cognitive and noncognitive users.

It is worth mentioning that, in general, depending on the availability  of side information  three types of behavior
can be defined for the  transmitters. (a) \textit{Competitive}: neither of the transmitters has
knowledge of the other transmitter's side information. (b) \textit{Cognitive}: only one transmitter (namely, the cognitive transmitter) has
knowledge of the other user's side information (see Figure~\ref{fig:cognitive}). (c) \textit{Cooperative}: both transmitters have
knowledge of the other user's side information. Throughout this chapter, we  focus on the cognitive behavior.

\subsection{Cognitive Radio Network Paradigms}
Depending on  the type of available network side information
and the regulatory constraints,  cognitive radio networks
can be divided into three main  paradigms \cite{goldsmith2009breaking}:
\textit{interweave}, \textit{underlay}, and \textit{overlay}.
While in the last two cases, the cognitive users concurrently transmit
over the same spectrum as the primary users, in the first case
cognitive users use \textit{spectrum holes} (temporary space-time frequency
voids) for transmission.

\begin{itemize}
   \item \textbf{Interweave (interference avoidance):} In the interweave
paradigm, cognitive users `opportunistically' use the spectrum so that their activity
does not  interfere the activity of noncognitive  users.
In other words, they only transmit during spectrum holes.
To avoid interfering with noncognitive  users, cognitive  users  require  knowing  the activity information of the
noncognitive  users in the shared spectrum. This paradigm, which is the simplest yet the most common paradigm
  was the original motivation for cognitive radio.
 \item \textbf{Underlay (interference control):} In this paradigm, the cognitive users can transmit
over the same spectrum as the noncognitive users provided that
the interference seen by the noncognitive users is
maintained to an acceptable level, i.e.,  certain QoS should be satisfied.
The cognitive users are often called \textit{secondary users} in this paradigm as
they are not allowed to significantly interfere with the communication
of  noncognitive (primary) users. Thus, they  require the knowledge of the ``acceptable levels'' of interference
at the primary users.
  \item \textbf{Overlay (interference mitigation):}
Similar to the underlay paradigm, in the overlay paradigm cognitive users can transmit simultaneously
with the noncognitive users. The main difference is
 that the cognitive users have the knowledge of the noncognitive users'  and possibly their messages
 codebooks in addition to their channel gains.
Thus, the cognitive users can allocate part of their power to relay the
noncognitive users' message. This can help boost the information rate at the noncognitive receivers.
On the other hand, the interference to cognitive users can be  mitigated or even canceled
by using  this side information (knowledge of  codebooks).
\end{itemize}

It is worth noting that  the first paradigm is also be referred to as \textit{opportunistic spectrum access} and
the other two paradigms may also be referred to as \textit{concurrent spectrum access} \cite{liang2011cognitive}.
Unless otherwise stated, in this chapter cognitive radio refers to overlay cognitive radio.

\subsection{Chapter Outline}
The chapter is structured as follows. In Section~\ref{sec:main},
we first define achievable rates  and capacity region for  the cognitive interference channel.
We then give a comprehensive summary of the
capacity  results established for this channel.
Our survey begins with the works on the simplest cognitive network,  i.e., a network
consisting of two pairs of transmitter-receiver, one cognitive and one noncognitive.
We will cover both  discrete memoryless and Gaussian channels. This will be
followed by  stating fundamental results for $K$-user and multi-antenna cognitive interference channels.
Our goal is not just to show how these capacity regions can be obtained but to get  intuition
into the optimal communication over this basic channel. In fact, rather than the capacity regions per se,
the techniques used to get such regions are important in this study. Such insight can be used to extend
the results to more complex networks. In Section~\ref{sec:5G},
we briefly describe the interplay cognitive radio  and emerging techniques in
wireless communication. This is followed by future research directions  in Section~\ref{sec:future}, which includes open problems.
We conclude the chapter in Section~\ref{sec:con}

\section{Cognitive Radio Channels: Capacity Results and Intuitions}\label{sec:main}

Information theory provides a framework  for analyzing the fundamental limits of
communication. Fundamental limits can then  be used as   benchmarks for the operation of
the desired communication system (cognitive radio networks here). This, in turns,
allows researchers and engineers to measure to what extent a practical network is efficient
and also  guides them in the design and standardization phases.


The two-user \textit{interference channel} (IC) is a two-transmitter two-receiver
network, in which each transmitter has an independent message for
its respective receiver \cite{carleial1975case,sato1981capacity, CostaIC, HK,vaezi2016simplified}.
The transmitters do not have side information about the other user's communication. Since users
 communicate over a shared channel, they interfere with each other. In
the cognitive radio communication setting, one transmitter
(cognitive transmitter) is able to sense the environment and
obtain side information about the other transmitter (noncognitive  or primary transmitter).
 Such a communication channel is called  {\it cognitive interference channel}, also known as interference channel
with ``unidirectional'' cooperation, or  simply  cognitive  channel.
We formally define this channel and its derivatives in the following.

\subsection{Discrete Memoryless  Channel}\label{sec:2userDMC}

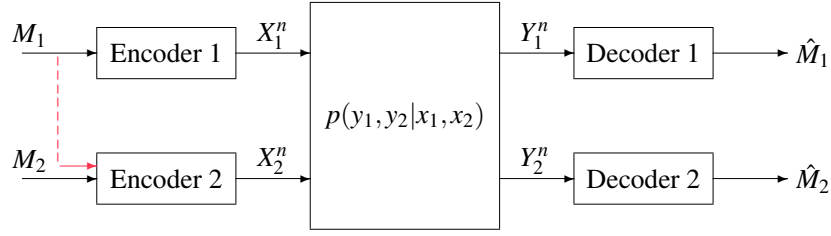
\begin{figure}
\begin{center}
\scalebox {0.95}{
\begin{picture}(240,100)

\put(-40,0){
\begin{picture}(200,50)
\put(20,20){\framebox(55,20){Encoder 2}}
\put(-10,30){\vector(1,0){30}}
\put(0,37){\makebox(0,0)[r]{$M_{2}$}}
\put(75,30){\vector(1,0){30}}
\put(96,37){\makebox(0,0)[r]{$X^n_{2}$}}
\put(105,10){\framebox(75,90){$p(y_1,y_2|x_1,x_2)$}}
\put(180,30){\vector(1,0){30}}
\put(200,37){\makebox(0,0)[r]{$Y^n_{2}$}}
\put(210,20){\framebox(55,20){Decoder 2}}
\put(265,30){\vector(1,0){30}}
\put(312,30){\makebox(0,0)[r]{$\hat{M}_{2}$}}
\end{picture}
}
\put(-40,50){
\begin{picture}(200,50)
\put(20,20){\framebox(55,20){Encoder 1}}
\put(-10,30){\vector(1,0){30}}
\put(0,37){\makebox(0,0)[r]{$M_{1}$}}
\put(75,30){\vector(1,0){30}}
\put(96,37){\makebox(0,0)[r]{$X^n_{1}$}}
\put(180,30){\vector(1,0){30}}
\put(200,37){\makebox(0,0)[r]{$Y^n_{1}$}}
\put(210,20){\framebox(55,20){Decoder 1}}
\put(265,30){\vector(1,0){30}}
\put(312,30){\makebox(0,0)[r]{$\hat{M}_{1}$}}
\put(2,5){{\color [rgb] {1,.2,0.3}\dashline}}
\put(5,-15){{\color [rgb] {1,.2,0.3}\vector(1,0){15}}}
\end{picture}
}
\end{picture}
}

\end{center}
\caption{The two-user discrete memoryless cognitive
interference channel (DM-CIC). Message $M_1$ is known to both Encoder~1 and Encoder~2, indicating that Encoder~2 corresponds to the cognitive user.  $M_2$ is known only to the cognitive encoder.
$X_1$ and $ X_2$ are the channel inputs, $Y_1$ and $ Y_2$ are the channel outputs, and $p(y_1,y_2|x_1,x_2)$ is the channel \textit{transition probability}.}
  \label{fig:DM-CIC}
\end{figure}

Consider a two-user  \textit{discrete memoryless} cognitive
interference channel (DM-CIC), depicted in Figure~\ref{fig:DM-CIC}, in which
user~$1$ and user~$2$  wish  to transmit independent messages   $M_1$ and $M_2$, respectively,  to their corresponding receivers.\footnote{We should highlight
that this  channel models the overlay paradigm, discussed earlier in this chapter.}
This channel is defined by a tuple $({\cal
X}_1,{\cal X}_2;p(y_1,y_2|x_1,x_2);{\cal Y}_1,{\cal Y}_2)$ where
${\cal X}_1,{\cal X}_2$ and ${\cal Y}_1,{\cal Y}_2$
are input and output alphabets and $p(y_1,y_2|x_1,x_2)$ is channel transition probability function.
A $(2^{nR_1}, 2^{nR_2}, n, \epsilon_1^n, \epsilon_2^n)$ code for this channel consists of
two independent messages $M_i$, $i\in\{1,2\}$,  two encoding functions $f_i$,  two decoding functions $g_i$, and two average probability
errors $\epsilon_i^n$, in which
\begin{enumerate}
  \item $M_i$  is uniformly distributed over  $[1,2, \hdots, 2^{nR_i}] $,
  \item encoder~$i$ assigns a codeword $x_i^n(m_i)$ to each message $m_i$
  \item decoder~$i$ assigns an estimate $\hat m_i \in [1,2, \hdots, 2^{nR_i}] $ to each received sequence $y_i^n$, and
  \item $\epsilon_i^n= p(\hat M_i \ne M_i )= \frac{1}{2^{nR_i}}\sum_{i=1}^{2^{nR_i}}p(\hat m_i \ne m_i )$.
\end{enumerate}

A rate pair $(R_1, R_2)$ is \textit{achievable} if there exist a sequence of codes
$(2^{nR_1}, 2^{nR_2}, n, \epsilon_1^n, \epsilon_2^n)$ with $\epsilon_1^n \to 0$ and $\epsilon_2^n \to 0.$
The \textit{capacity region} of this channel is the closure of the  set of achievable rates.

Motivated by cognitive radio's promise to increase the spectral
efficiency in wireless systems, the study of interference channel
with cognitive users has been receiving increasing attention during the past years.
Fundamental limits of the cognitive
interference channel, in which the cognitive transmitter non-causally
knows the full message of the primary user has been studied in
\cite{Devroye,Jovicic-Viswanath,Wu-Vishwanath,Maric2,Maric3,rini2011new,vaezi2011capacity,JiangZ,vaezi2011superposition,rini2012inner}.
This channel was first introduced in \cite{Devroye} where the authors obtained achievable rates by applying
Gel'fand-Pinsker coding \cite{GP} to the
celebrated Han-Kobayashi encoding \cite{HK} for the
IC. The capacity of this channel remains unknown in general; however, it is known
in several special cases, both in the discrete memoryless and Gaussian channels.

\begin{table}[tb]
\renewcommand{\arraystretch}{.9}
\caption{The summary of capacity results for the DM-CIC.} \vspace{.15cm} \label{table1}
\centering
\scalebox{.8}{
\begin{tabular}{|c|c|c|c|c|}
\hline
\bfseries Label &\bfseries DM-CIC class &\bfseries Condition & \bfseries Capacity region &  \bfseries Ref. \\
\hline\hline\

$\mathcal C_I$ & \mbox {cognitive-less-noisy} & $ I(U;Y_1) \leq I(U;Y_2) $ & $ R_1 \leq I(U;Y_1) $  & \cite{vaezi2012lessnoisy} \\
$  $ &  & $ $ & $ R_2 \leq I(X_2;Y_2|U) $   &  \\
\hline

$\mathcal C_{II}$  & \mbox {strong interference} &  $I(X_1,X_2;Y_1) \leq I(X_1,X_2;Y_2)$ & $  R_1  \leq I(X_1;Y_1) $  &  \cite{Maric2} \\
$ $ & & $ I(X_2;Y_2|X_1) \leq I(X_2;Y_1|X_1) $ & $ R_2 \leq I(X_2;Y_2|X_1) $   &  \\
\hline

$\mathcal C_{III}$ & \mbox {weak interference} & $ I(X_1;Y_1) \leq I(X_1;Y_2) $ & $ R_1 \leq I(U,X_1;Y_1) $ & \cite{Wu-Vishwanath} \\
 $ $ & & $ I(U;Y_1|X_1) \leq I(U;Y_2|X_1)  $ & $ R_2 \leq I(X_2;Y_2|U,X_1) $  &  \\
\hline

$$  & & $ $ & $ R_1 \leq I(U,X_1;Y_1) $   &  \\
$ \mathcal C_{III}^\prime$ & \mbox {better-cognitive-decoding} & $  I(U,X_1;Y_1) \leq I(U,X_1;Y_2) $ & $ R_2 \leq I(X_2;Y_2|X_1) $ & \cite{rini2011new} \\
$ $ & & $ $ & $ R_1 + R_2 \leq I(U,X_1;Y_1) + I(X_2;Y_2|U,X_1) $ &   \\
\hline

$$  & & $ $ & $ R_1 \leq I(U,X_1;Y_1) $   &  \\
$ \mathcal C_{IV}$ & \mbox {cognitive-more-capable} & $  I(X_1,X_2;Y_1) \leq I(X_1,X_2;Y_2) $ & $ R_2 \leq I(X_2;Y_2|X_1) $   & \cite{vaezi2014more} \\
$ $ & & $ $ & $ R_1 + R_2 \leq I(U,X_1;Y_1) + I(X_2;Y_2|U,X_1)$ &  \cite{farsani2015capacity}
\\$ $ & & $ $ & $ R_1 + R_2 \leq I(X_1, X_2;Y_2) $ &   \\
\hline

\end{tabular}}

\end{table}

\begin{figure}[!tb]
\begin{center}
\scalebox {1}{
\begin{tikzpicture}
\draw[black, thick, rotate=0] (-4,2.7) rectangle (4,-2.7);
\draw[blue, thick, rotate=0] (0,0) ellipse (110pt and 65pt);
\draw[red, dashed, thick, scale=1.5,rotate=0] (0,0)ellipse (58pt and 32pt);
\draw[black, dotted, thick, scale=1.5,rotate=0] (0,0)ellipse (33pt and 8pt);
\draw[gray, densely dotted, thick, scale=1.5,rotate=0] (0,0)ellipse (45pt and 18pt);
\draw  (0,2.8)node[right, above] {DM-CIC};
\draw  (0,1.6)node[right, above] {{\color[rgb]{.1,0.1,1} cognitive-more-capable}};
\draw  (0,.85)node[right, above]  {{\color[rgb]{1,0.1,.1} better-cognitive-decoding}};
\draw  (0,-0.25)node[right, above] { cognitive-less-noisy} ;
\draw  (0,.35)node[right, above] {{\color[rgb]{.5,.5,.5} strong interference}} ;
\end{tikzpicture}
}
\end{center}
\caption{The class of the DM-CIC.
The cognitive receiver is {\it superior} to the primary receiver for the cognitive-more-capable and all its subclasses.
 }
\label{fig:CICclass}
\end{figure}
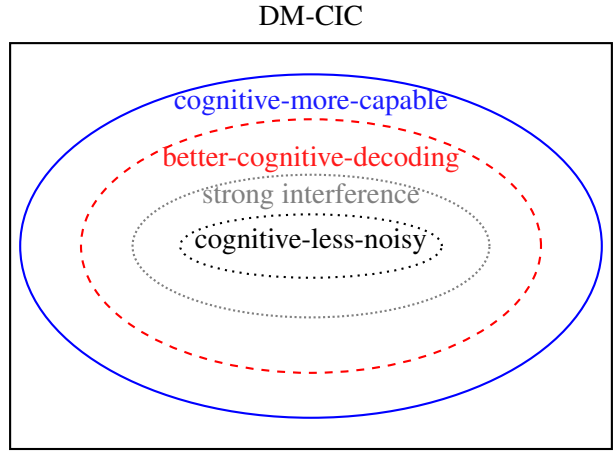

The capacity of the DM-CIC is known for several classes,  including the cases in which the cognitive user
is less noisy or more capable than the primary user,
as well as weak and  strong interference regimes. These capacity regions and their corresponding  conditions are listed in Table~\ref{table1}.
 It can be checked that $\mathcal C_I   \subseteq \mathcal C_{II} \subseteq \mathcal C_{III}  \subseteq \mathcal C_{IV}$  \cite{vaezi2014more}
 and $\mathcal C_{III}^\prime \equiv \mathcal C_{III}$ \cite{vaezi2012comments}
 For all of the above cases,
the cognitive receiver has a better condition (more information) than the primary one in some sense,
as it can be understood from the corresponding conditions in Table~\ref{table1}. It is important to note that the
\textit{cognitive-more-capable} channel (labeled $\mathcal C_{IV}$) includes all other cases as its subcases (see Fig.~\ref{fig:CICclass} and \cite{vaezi2014more}).
For this reason, it suffices to discuss the achievability scheme  for this case.

The achievability scheme of the capacity region of the cognitive-more-capable is based on \textit{superposition coding} at the cognitive transmitter.
With sophisticated schemes, which combine other techniques such as \textit{rate-splitting}
and \textit{Gel'fand-Pinsker coding (binning)} with superposition coding, one may enlarge the
 achievable rate region when $  I(X_1,X_2;Y_1) \nleq I(X_1,X_2;Y_2) $
 \cite[Theorem~7]{rini2011new}. However, it is not clear how much gain this complication brings in.
 In addition, such techniques (e.g., binning) are too  complicated
 to be used in practical networks.

It is worth pointing out that the capacity region of the cognitive-more-capable DM-CIC, given in Table~\ref{table1},  is the same
as the capacity region of the DM-CIC in which the cognitive receiver (Receiver~2) needs to decode both messages.
The capacity region in the  latter case is obtained in  \cite[Theorem 4]{liang2009capacity}. Interestingly, this additional constraint, i.e., the constraint that
the cognitive receiver must also decode $M_2$,  leads to the determination of the capacity region of the DM-CIC for any channel condition.
On the contrary,  the cognitive-more-capable DM-CIC by definition implies $ I(X_1,X_2;Y_1) \leq I(X_1,X_2;Y_2) $; that is, the capacity region of this channel
is valid only if the aforementioned  condition on the channel  holds. Comparing the two capacity results, we conclude that in the cognitive-more-capable DM-CIC
channel the cognitive receiver can decode both messages.

\begin{rem}
\textit{Superposition coding} at the cognitive transmitter is the capacity-achieving technique in all above cases.
 Nonetheless, more complicated techniques, such as rate splitting and banning, are reported to result in
 a larger achievable region, in general.
 %
 \end{rem}

\begin{rem}
 When the cognitive transmitter and receiver are in the vicinity of the noncognitive transmitter
and far away from the noncognitive receiver, there is a high possibility for
the cognitive receiver to be more capable the noncognitive receiver; i.e., $ I(X_1,X_2;Y_2)\ge I(X_1,X_2;Y_1) $  hold.
As discussed earlier, in such a case,  superposition coding at the cognitive transmitter is optimal.
\end{rem}

\subsection{Gaussian  Channel}\label{sec:2userG}

In this subsection, we study the two-user Gaussian cognitive interference channel (GCIC). We first
 describe the channel model  and then summarize the previously known results for the GCIC
as well as the one-sided GCIC.

\subsubsection{Two-User Gaussian  Channel}
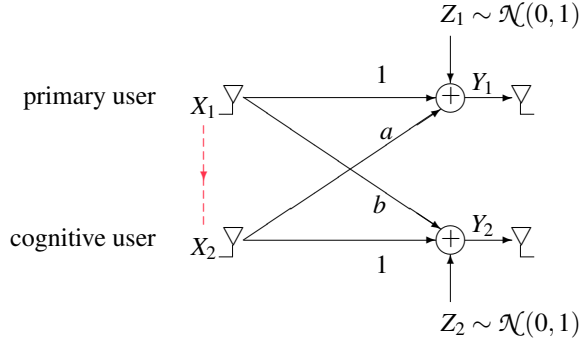
\begin{figure}{t}
\begin{center}
\scalebox{.9}{
\begin{picture}(100,180)
\put(-20,50){
\begin{picture}(200,80)
\put(96,10){\circle{12}}
\put(92,8){$+$}
\put(9,10){\vector(1,0){81}}
\put(102,10){\vector(1,0){20}}
\put(96,-16){\vector(0,1){20}}
\put(9,10){\vector(3,2){83}}

\put(5,10){\makebox(-7,-7)[r]{$ X_2$}}
\put(-25,10){\makebox(0,0)[r]{cognitive user }}
\put(0,10){$\bigtriangledown$}
\put(4.45,3){\line(0,1){5}}
\put(-1.35,3){\line(1,0){6}}

\put(70,0){\makebox(0,0)[r]{$1$}}
\put(69,25){\makebox(0,0)[r]{$b$}}
\put(105,16){\makebox(0,0)[l]{$Y_2$}}
\put(92,-25){\makebox(0,0)[l]{$Z_2 \sim {\cal N} (0, 1)$}}

\put(121,10){$\bigtriangledown$}
\put(125.45,3){\line(0,1){5}}
\put(125.25,3){\line(1,0){6}}
\end{picture}
}

\put(-20,100){
\begin{picture}(200,80)
\put(96,20){\circle{12}}
\put(92,18){$+$}
\put(9,20){\vector(1,0){81}}
\put(102,20){\vector(1,0){20}}
\put(96,46){\vector(0,-1){20}}
\put(9,20){\vector(3,-2){83}}
\put(-10,-15){\color [rgb] {1,.2,0.3}\dashline}
\put(-7.6,-15){{\color [rgb] {1,.2,0.3}\vector(0,-2){0}}}

\put(5,10){\makebox(-7,12)[r]{$ X_1$}}
\put(0,20){$\bigtriangledown$}
\put(4.45,13){\line(0,1){5}}
\put(-1.35,13){\line(1,0){6}}

\put(-25,20){\makebox(0,0)[r]{primary user }}

\put(70,30){\makebox(0,0)[r]{$1$}}
\put(72,5){\makebox(0,0)[r]{$a$}}
\put(105,26){\makebox(0,0)[l]{$Y_1$}}
\put(92,55){\makebox(0,0)[l]{$Z_1 \sim {\cal N} (0, 1)$}}

\put(121,20){$\bigtriangledown$}
\put(125.45,13){\line(0,1){5}}
\put(125.25,13){\line(1,0){6}}
\end{picture}
}
\end{picture}
}
\end{center}
\vspace{-20pt}
\caption{ A Gaussian cognitive interference channel in the standard form, with inputs $ X_{1}$ and
 $ X_{2}$, outputs $ Y_{1}$ and  $Y_{2}$,  noises $ Z_{1}$ and
  $Z_{2}$, and interference gains $a$ and $b$. }
  \label{fig:GCIC}
\end{figure}

The two-user Gaussian cognitive interference channel, depicted in   Figure~\ref{fig:GCIC},
 is composed of  two transmitter-receiver pairs
in which each transmitter communicates with its corresponding
receivers while interfering with the other receiver. This model is very similar to that of the
 two-user Gaussian interference channel; the only difference is in that the cognitive transmitter knows
 the message (and possibly the codewords) of the primary user. This flow of information is shown by the dashed line
 in Figure~\ref{fig:GCIC}.

Without loss of generality, we use the {\it standard form} of the Gaussian
interference channel \cite{kramer2006review}, in which, for a single channel use,  the
 channel is  expressed   by
\begin{subequations}\label{defIC}
\begin{align}
Y_1 &= X_1 + aX_2 + Z_1, \label{defIC:first}\\
Y_2 &= bX_1 + X_2 +  Z_2,\label{defIC:second}
\end{align}
\end{subequations}
where   $a$ and $b$ are  two non-negative real numbers
representing the crossover gains; and, for $j \in \{1, 2\}$, $ X_{j}$, $Y_{j}$, and $Z_{j}$, respectively, represent
the  transmitted signal, received signal, and the channel noise, and
$Z_1$ and $ Z_2$ are independent and identically distributed (i.i.d.)  Gaussian random variables with zero means and unit
variances. Let $M_1$ and $M_2$ be two independent messages  uniformly distributed
over $\mathcal{M}_1=[1,\hdots,2^{nR_1}]$ and $\mathcal{M}_2=[1,\hdots,2^{nR_2}]$, respectively.\footnote{For $j \in \{1, 2\}$, $M_j$ is a random variable
distributed over set $\mathcal{M}_j$, and $m_j$ is a realization of $M_j$.}
Transmitter $j$ wishes to transmit message $M_j$
to receiver $j$ in $n$ channel uses at rate $R_j$, and $X_{j}$ is subject to an average
power constraint $P_{j}$, i.e.,
\begin{align}
\frac{1}{n}\sum_{i=1}^{n}\|X_{ji}\|^2 \leq P_{j}, \;j=1,2.
\label{eq:power}
\end{align}
The capacity region of this channel is defined as the set of all
rate pairs $(R_1, R_{2})$ for which each receiver is able to
decode its own message with arbitrarily small probability of error.
Capacity region of the Gaussian cognitive interference channel is
known at certain interference regimes.
All capacity regions know for this channel are based on using a combination of
\textit{dirty paper coding} (DPC) \cite{Costa} and \textit{superposition coding}
at the cognitive user. Before stating the capacity region, we discuss this
achievable region in the following.

As discussed earlier, in the cognitive interference channel, the cognitive user
knows the noncognitive user's messages and codewords. This signifies that the
cognitive user can use this knowledge to cancel the interference received from
the noncognitive user via DPC. On the other hand,  to compensate
for the interference the noncognitive transmitters creates on
 the cognitive receiver and, thus,  to improve the achievable rate at the
 noncognitive receiver,  it would be useful if the cognitive user allots part of its power to help send the
codewords of the primary user. The latter scenario implies  superposition coding.

Not surprisingly, an optimal encoding strategy at the cognitive transmitter is to use
DPC to encode $M_2$ while treating $X_1$ as interference and, then,  superimpose  $M_1$ on top of that to help convey $M_1$ to Receiver~1.
Superposition coding implies that the cognitive user partially uses its power to help send the
codewords of the primary user. $X_2$ contains two independent Gaussian parts,
$X_2 = \sqrt{\alpha P_2}V_1(m_1)+\sqrt{\bar{\alpha} P_2}V_2(m_2)$, in which $V_1$ and $V_2$ are
auxiliary random variables used to encode $m_1$ and $m_2$, respectively,
and $0\le \alpha \le 1$ and $\bar \alpha = 1- \alpha$.
The primary user, however, does not have a knowledge about the cognitive user's messages;
thus, it uses its whole power to transmit $m_1$, i.e.,
$X_1 = \sqrt{P_1}V_1(m_1)$.

For decoding, one strategy is to let  the noncognitive receiver (Receiver~1) simply decodes its
own codeword assuming the other codeword as interference. From \eqref{defIC:first}, it is seen that
$Y_1 = \sqrt{P_1}V_1(m_1) + a \sqrt{\alpha P_2}V_1(m_1) +a \sqrt{\bar \alpha P_2}V_2(m_2) + Z_1$. This indicates that
$(\sqrt{P_1}  + a \sqrt{\alpha P_2})V_1(m_1)$ is the useful signal at Receiver~1 while $a \sqrt{\bar \alpha P_2}V_2(m_2)$ is
the interference. Therefore,  $R_{1} \leq \frac{1}{2}\log{\big(1 + \frac{(\sqrt{P_1} + |a|
       \sqrt{\alpha  P_2 })^2} {1+ a^2 \bar{\alpha} P_2}}\big)$ is achievable  by treating the interference as noise.
On the other hand, in view of \eqref{defIC:second}, the signal seen by the cognitive receiver can be expressed
as $Y_2 = (b\sqrt{P_1} + \sqrt{\alpha P_2})V_1(m_1)+\sqrt{\bar{\alpha} P_2}V_2(m_2)+  Z_2.$
Due to the DPC at the cognitive transmitter, the cognitive receiver  can cancel the interference
 $V_1(m_1)$; thus, $ R_2 \leq \frac{1}{2}\log{(1+\bar{\alpha} P_2)} $ is achievable.
Finally, considering the error analysis for the sum rate, the above encoding and decoding result in
the following achievable  rate region \cite{Jovicic-Viswanath}: 

\begin{lem}\label{lem1}
The  set of rate pairs  $(R_1, R_{2})$ satisfying
\begin{subequations}\label{eq:A}
\begin{align}
R_{1} &\leq \frac{1}{2}\log{\Big(1 + \frac{(\sqrt{P_1} + |a|\sqrt{\alpha  P_2 })^2} {1+ a^2 \bar{\alpha} P_2}}\Big), \label{eq:A1} \\
R_2 &\leq \frac{1}{2}\log{(1+\bar{\alpha} P_2)}, \label{eq:A2}\\
R_1 + R_2 &\leq \frac{1}{2}\log{ \Big(1+P_1 + a^2 P_2 + 2|a| \sqrt{\alpha P_1P_2}  \Big) }, \label{eq:A3}
\end{align}
\end{subequations}
in which $0 \le \alpha \le 1$ and  $\bar \alpha = 1 -\alpha$ is achievable for the cognitive interference channel.
\end{lem}

%
%
%
%
%
%

The above rate region simplifies to the capacity region of the cognitive interference channel under
certain channel conditions, as listed below.

\begin{itemize}
  \item\textbf{Weak interference ($|a| \le 1$)} \cite{Jovicic-Viswanath,Wu-Vishwanath}: In this regime,  the
  optimal encoding strategy at the cognitive transmitter is to use
DPC and superposition coding, as explained in the achievability of the above rate region.
In particular, since the interference channel gain is small,  Receiver~1 does not attempt to
decode the interference. It simply treats interference as noise and this turns out to be the optimal solution.
Moreover, it can be checked that \eqref{eq:A3} is redundant for $|a| \le 1$ and the capacity region is obtained by
\begin{subequations}
\begin{align}
R_{1} &\leq \frac{1}{2}\log{\Big(1 + \frac{(\sqrt{P_1} + |a|\sqrt{\alpha  P_2 })^2} {1+ a^2 \bar{\alpha} P_2}}\Big), \\
R_2 &\leq \frac{1}{2}\log{(1+\bar{\alpha} P_2)}.
\end{align}
\end{subequations}

  \item \textbf{Strong interference ($|a| > 1$):}  In this case, since $|a| > 1$,
  the interference at Receiver~1 is stronger than that
in the weak interference case. As a result, depending on the value of $|a|$, decoding $M_1$,
 or a part of the interference (unwanted message), can  be beneficial.
In \cite[Theorem~6]{Maric2} it is proved that both users can decode both messages when
$|a| \ge 1 $, $ |b\gamma-1| \ge |a-\gamma| $,  and  $ |b\gamma+1| \ge |a+\gamma| $
where $ \gamma \triangleq \sqrt{P_1/P_2}$.\footnote{In this case,
the channel becomes a  compound multiple access channel (MAC) and the capacity region of compound MAC is applicable. }
In such a case
\begin{subequations}\label{eq:Aa>1}
\begin{align}
R_2 &\leq \frac{1}{2}\log{(1+\bar{\alpha} P_2)}, \label{eq:A2a>1}\\
R_1 + R_2 &\leq \frac{1}{2}\log{ \Big(1+P_1 + a^2 P_2 + 2|a| \sqrt{\alpha P_1P_2}  \Big) }, \label{eq:A3a>1}
\end{align}
\end{subequations}
 characterize the capacity region.

 In \cite{rini2012inner}, it is shown that  the encoding and decoding strategy resulting Lemma~\ref{lem1} can be optimal when
$|a| \ge 1 $. Specifically, it is shown that the above
 inequalities  also give the capacity region when
 $|a| \ge 1 $ and   $ P_1|1 - |a|b|^2 \ge (|a|^2 -1)(1 + P_2 + |b|^2 P_1 ) - P_1P_2|1 - |a|b|^2$.
 Noting that  \eqref{eq:A1} is redundant for $|a| \ge 1 $, we can see that \eqref{eq:Aa>1}
 is also the capacity region in this case. 
  The above two set of conditions  
 be both valid under certain channel realizations. This indicates that  more than one scheme can be optimal
at least at certain channel conditions. Specifically, in the above cases the interference is canceled in two
radically different ways, i.e, by decoding and then canceling it versus  using DPC. The former does not require any
information at the encoder while the latter requires knowing the interference at the encoder and applies
a very complex encoding.

 \item \textbf{Cognitive receiver needs to decode both  messages:}
 The capacity region of the Gaussian cognitive interference channel is also known
when the cognitive receiver needs to decode both users' message \cite{liang2009capacity}.
\end{itemize}

\subsection{Gaussian Z-Channel}\label{sec:2userZ}

A  Z-Channel (or one-sided interference channel) models a  two-transmit two-receiver
scenario in which one of the users
does not experience interference. In a cognitive channel, due to asymmetric transmitters in which only
one transmitter has information about the other, two different ZICs are
conceivable: one  with no interference at the noncognitive receiver ($a=0$)
 and the other one with no interference at the cognitive receiver ($b=0$).
The capacity region of the former case is a special case
of the capacity region of the cognitive  interference channel
in the weak interference regime($|a| \le 1$),
and is obtained by DPC at the noncognitive transmitter  \cite{Jovicic-Viswanath} and \cite{Wu-Vishwanath}.
In the latter case, the capacity region is open in general. However, it is known in several special
cases, as discussed in the following.

\begin{figure}
\begin{center}
\scalebox{.8}{
\begin{picture}(100,180)
\put(0,50){
\begin{picture}(200,80)
\put(96,10){\circle{12}}
\put(92,8){$+$}
\put(9,10){\vector(1,0){81}}
\put(102,10){\vector(1,0){20}}
\put(96,-16){\vector(0,1){20}}
\put(9,10){\vector(3,2){83}}

\put(5,10){\makebox(-7,-7)[r]{$ X_2$}}
\put(0,10){$\bigtriangledown$}
\put(4.45,3){\line(0,1){5}}
\put(-1.35,3){\line(1,0){6}}

\put(70,0){\makebox(0,0)[r]{$1$}}
\put(105,16){\makebox(0,0)[l]{$Y_2$}}
\put(92,-25){\makebox(0,0)[l]{$Z_2 \sim {\cal N} (0, 1)$}}

\put(121,10){$\bigtriangledown$}
\put(125.45,3){\line(0,1){5}}
\put(125.25,3){\line(1,0){6}}
\end{picture}
}

\put(0,100){
\begin{picture}(200,80)
\put(96,20){\circle{12}}
\put(92,18){$+$}
\put(9,20){\vector(1,0){81}}
\put(102,20){\vector(1,0){20}}
\put(96,46){\vector(0,-1){20}}

\put(5,10){\makebox(-7,12)[r]{$ X_1$}}
\put(0,20){$\bigtriangledown$}
\put(4.45,13){\line(0,1){5}}
\put(-1.35,13){\line(1,0){6}}
\put(-10,-15){\color [rgb] {1,.2,0.3}\dashline}
\put(-7.6,-15){{\color [rgb] {1,.2,0.3}\vector(0,-2){0}}}


\put(70,30){\makebox(0,0)[r]{$1$}}
\put(50,-8){\makebox(0,0)[r]{$\sqrt{a}$}}
\put(105,26){\makebox(0,0)[l]{$Y_1$}}
\put(92,55){\makebox(0,0)[l]{$Z_1 \sim {\cal N} (0, 1)$}}

\put(121,20){$\bigtriangledown$}
\put(125.45,13){\line(0,1){5}}
\put(125.25,13){\line(1,0){6}}
\end{picture}
}
\end{picture}
}
\end{center}
\vspace{-15pt}
\caption{A one-sided Gaussian interference channel in  standard form.}
\label{fig:StandardOSIC}
\end{figure}
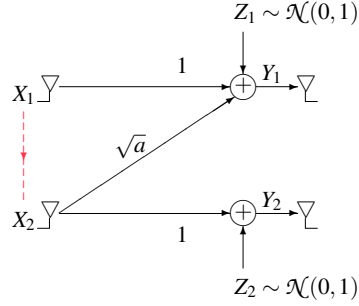

Consider a two-user cognitive Gaussian interference channel in which $b=0$, as shown in Figure~\ref{fig:StandardOSIC}.
The capacity region of this channel  is established in several ranges of interference gain
\cite{vaezi2011capacity,JiangZ,rini2012inner,vaezi2011superposition};
these results are summarized in Table~\ref{table3}.
 While in the low interference regime a
combination of dirty paper coding and superposition coding is the capacity-achieving scheme, in the high
interference regime superposition coding single-handedly can achieve the capacity region.
From this table, it is clear that
the capacity region of the cognitive Z-Channel is unknown only when $ \sqrt { 1 + P_1/(1+P_2) }< |a| < \sqrt{1 + P_1 }  $.

It is known that time-sharing can increase the  achievable rates for the interference channel and one-sided interference
channel  \cite{HK,vaezi2015limiting}.
Similarly, time-sharing can increase the secrecy capacity of the IC, see \cite[Lemma 3]{bustin2016secrecy}, for example.
It would be interesting to apply time-sharing to the secrecy capacity of cognitive interference channel.

\begin{table}[!t]
\renewcommand{\arraystretch}{1.3}
\caption{The summary of  capacity results for the GCZIC} \label{table3}
\centering
\scalebox{.8}{
\begin{tabular}{|c|c|c|c|}
\hline
\bfseries Condition & \bfseries Capacity region & \bfseries Technique  & \bfseries Ref. \\
\hline\hline\
 $|a| \leq 1$ & $
   R_{1} \leq \frac{1}{2}\log{\Big(1 + \frac{(\sqrt{P_1} + |a|
       \sqrt{\alpha  P_2 })^2} {1+ a^2 \bar{\alpha} P_2}}\Big)
 $ & superposition coding  & \cite{Jovicic-Viswanath},  \\
 $  $ & $ R_2 \leq \frac{1}{2}\log{(1+\bar{\alpha} P_2)} $ & and DPC   &  \cite{Wu-Vishwanath}\\
\hline

 $1 \leq |a| \leq  \sqrt {1 + \frac{P_1}{1+P_2}}$ & $
   R_1 + R_2 \leq \frac{1}{2}\log{ \Big(1+P_1 + a^2 P_2 + 2|a| \sqrt{\alpha P_1P_2}
   \Big) } $ & superposition coding &  \cite{JiangZ}, \\
 $  $ & $ R_2 \leq \frac{1}{2}\log{(1+\bar{\alpha} P_2)} $ &  and DPC &  \cite{rini2012inner} \\
\hline

$ \sqrt { 1 + \frac{P_1}{1+P_2}} < |a| < \sqrt{1 + P_1 }  $ & unknown & unknown  & ---\\
\hline

$  $ & $  R_1\leq \frac{1}{2}\log{ \left (1 + (\sqrt{P_1} + |a| \sqrt{\alpha P_2 })^2 \right)}
$  & & \\
 $ |a| \geq \sqrt {1 + P_1} $ & $  R_2 \leq \frac{1}{2}\log{\Big(1 +
    \frac{\bar{\alpha}P_2}{1+\alpha P_2}\Big)}$ & superposition coding  &  \cite{vaezi2011superposition}   \\
$  $ & $ R_1 + R_2 \leq \frac{1}{2}\log{ \Big(1+P_1 + a^2 P_2 + 2|a| \sqrt{\alpha P_1P_2}
   \Big) }$ &  &  \\
\hline
%
\end{tabular}}
\end{table}

\subsection{Capacity Approximation}\label{sec:2userG}
Finding the exact capacity region for many multi-user channel networks
has appeared to be daunting challenging. Considering this difficulty, one way to get insights
into the behavior of different multi-user channels  is to resort to approximation. Two
approximation metrics have gained significant attention during the past decade.
These are \textit{degrees of freedom} (DoF) and \textit{generalized degrees of freedom} (GDoF)
respectively.

\begin{itemize}
  \item\textbf{degrees of freedom (DoF):}
  The DoF or the \textit{multiplexing gain} is a means of approximating
   the sum capacity of a channel/network.\footnote{DoF region is a similar metric which studies both individual and sum rates.}
   It gives the \textit{pre-log} of the sum-rate capacity of a given multi-user channel in the high SNR regime.
Although rather coarse, DoF provides an analytically tractable way to characterize
the sum capacity in a given multi-user channel in the high SNR regime.
For example, the DoF for $K$-user Gaussian interference channel is shown to be  $\frac{K}{2}$,
and can be achievable through the interference alignment (IA).
This means that each user can enjoy half of the spectrum in the high SNR regime.

  \item\textbf{generalized degrees of freedom (GDoF):}
The GDoF  generalizes the notion of the DoF into different SNR regimes
and, thus, is a much more powerful metric. The GDoF is also known as the capacity region to within a
constant gap.
The insight obtained for the DoF  may not hold true for the GDoF. As an important example,
it is known that the relay does not increase
the DoF of the interference channel with relay whereas it can  increase the GDoF of
that channel \cite{chaaban2012generalized}.
\end{itemize}

While the DoF and GDoF approximate the sum-capacity,
there are also metrics to determine  either an \textit{additive}  or a  \textit{multiplicative} gap  between the inner
and outer bounds for a certain channel, rather than only  their sum-capacity. An additive gap between the inner
and outer bounds is useful at high signal-to-noise  power ratios (SNR)
 because in such a regime the difference
between inner and outer bound is small in comparison
to the magnitude of the capacity region. A multiplicative
gap is useful at low SNR, where the ratio between the inner
and outer bounds can be a better indicator  of their distance.

Etkin et al. obtained an approximation of the capacity region of the
  real-valued two-user Gaussian interference channel to within $\frac{1}{2}$ bits in \cite{etkin2008gaussian}.
  Rini et al. \cite{rini2012inner}, found the capacity region of the two-user
real-valued Gaussian cognitive interference channel to within  1.87 bits/s/Hz. This
 constant gap was obtained by using insights from the high
SNR deterministic approximation of the Gaussian cognitive interference
channel. Additive gap on the capacity region of this channel is known to within
1.87 bit/s/Hz \cite{rini2012inner} while the multiplicative gap is known to within  2 bits/s/Hz.
To achieve the multiplicative gap one can use a simple \textit{time-sharing} between the following
two achievable points:
\begin{align}
A = (R_1^A, R_2^A)&= (\frac{1}{2}\log{ \left (1 + (\sqrt{P_1} + |a| \sqrt{ P_2 })^2 \right)}, 0),\\
B = (R_1^B, R_2^B)&= (0, \frac{1}{2}\log{ \left (1 + P_2 \right)}).
\end{align}
It can be seen that to achieve the point $A$ the cognitive user sacrifice its
rate and only transmits the codewords of the noncognitive user.
On the other hand, to achieve the point $B$  the noncognitive user must be silent
while the cognitive user transmits only its own codewords.
Finally, for different values of $a$, the additive gap is obtained by applying different achievable schemes
 in \cite[Table II]{rini2012inner}.

\subsection{Further Capacity Results}\label{sec:fur}
\subsubsection{Secrecy Capacity}\label{sec:2userG}
The secrecy capacity of two-user DM-CIC is studied in \cite{liang2009capacity,liu2008discrete,bafghi2010cognitive}. In \cite{liang2009capacity}
 it is assumed that $M_2$ is confidential and  needs to
be kept  secret from noncognitive
receiver  (Rx1 in Fig~\ref{fig:DM-CIC}); in addition,
it is assumed that the cognitive receiver decodes both messages whereas the noncognitive
receiver  decodes only message 1.
 This is different from \cite{liu2008discrete} in that it is not assumed that the cognitive receiver
 decodes both messages; it also does not assume that the  cognitive transmitter  knows
the other user's message. The DM-CIC
 with two confidential messages is studied in  \cite{bafghi2010cognitive}, in which both  primary and cognitive messages must be secure at unintended receivers.

\subsubsection{Multi-User Channels}\label{sec:2userG}
So far we have  focused on the two-user channels which include one noncognitive and one cognitive user.
In general, multiple cognitive and multiple
noncognitive users in the overlay network can simultaneously share the same spectrum.
The extension of the capacity results of the  two-user channels to \textit{3-use}r and,  in general,
\textit{$\mathrm{K}$-user} channels is not straightforward. To find fundamental limits of these channels,
the techniques  used for establishing the capacity results in the previous sections can be used.
For example, the rate-splitting approach can be generalized as a way to cope with interference from multiple senders.
However, such a scheme becomes extremely complicated when the number of users increases.
Interference alignment can be promising approaches for the $\mathrm{K}$-user interference channel.
Interference alignment in cognitive nodes can
reduce the interference at both the noncognitive and the cognitive receivers.
Recall that cognitive users can  perform relaying
of noncognitive messages and precoding against interference. Understanding the  interplay between these
techniques is  an important and interesting research topic. A survey on multi-user cognitive interference channels
can be found in \cite{maamari2015multi}.

The capacity region of the \textit{multicast} cognitive interference channel
in which each transmitter   wishes to transmit an independent message
 to a  set of users is investigated \cite{benammar2017capacity}.
 This channel can be seen as a two-user cognitive interference channel in
 which user~$1$ and user~$2$  wish  to transmit independent messages   $M_1$ and $M_2$, respectively,  to  $Y_{11}, \hdots,Y_{1N_1}$
 and $Y_{11}, \hdots,Y_{1N_2}$, where $N_1\ge 1$ and $N_2\ge 1$ are arbitrary integers.
 The paper has interesting capacity results for multi-primary ($N_1\ge 2$ and $N_2 =1$)
  and multi-secondary ($N_1=1$ and $N_2 \ge 2$)  cognitive interference channels in various interference regimes,
  including very strong, very weak, and mixed very weak/strong interference regimes.
These capacity results are mainly the extensions of the capacity results in
\cite{Jovicic-Viswanath,Wu-Vishwanath} and are a step forward toward the
scenarios where multiple users wish to communicate over the same chunk of spectrum.

\subsubsection{MIMO Channel}\label{sec:mimo}

multiple-input and multiple-output (MIMO) communication can also be exploited in the cognitive radio networks as a potential method for the spectrum sharing.
Multiple antenna techniques  can be used for throughput enhancement and interference cancellation. Fundamental limits of MIMO cognitive radio has been
studied in the literature. Most of the results,  however, discuss  the MIMO cognitive interference channel from the DoF perspective,  either with perfect or delayed channel
state information at transmitter (CSIT) \cite{huang2009degrees,vaze2012degree}.

It is known that cognitive message sharing can increase the sum DOF of the MIMO cognitive interference
channel for certain scenarios. Further, in terms of sum DOF,
having  a cognitive transmitter is more beneficial than  having a cognitive receiver.
Specifically, for a MIMO Gaussian interference channel with $L_1, L_2$ antennas at transmitters and
$N_1, N_2$ antennas at receivers the following DoF results are obtained in \cite{huang2009degrees}.

\begin{itemize}
  \item\textbf{cognitive message sharing:}
  For the case of cognitive message sharing in which only the transmitter of the secondary
user (transmitter~2) knows the message of transmitter~1, the sum DOF is
  given by
  \begin{align}
  \min\{L_1+L_2,\; N_1+N_2,\;   \max(L_2, N_1) \}.
    \end{align}
 Note that this is  an information-theoretic
setting where  the message of the
primary user is provided by a \textit{genie} to the transmitter of the secondary
user  \textit{noncausally} and  \textit{without noise}.
  \item\textbf{cooperation at transmitters:}
 User cooperation  refers to the case where several distributed nodes can cooperate with each other to form a transmit antenna array or a receive
antenna array. The links between cooperating  transmitters or cooperating  receivers are assumed to be
\textit{noisy}.
For the case of users'  cooperation (be it at the transmitters side, receivers side, or both sides),  the sum DOF is
   \begin{align}\label{eq:sumDoFIC}
   \min\{L_1+L_2,\; N_1+N_2,\;   \max(L_1, N_2),\;  \max(L_2, N_1) \}.
     \end{align}

\end{itemize}

Note that \eqref{eq:sumDoFIC}  is the same as the sum DOF of the channel without cooperation \cite{huang2009degrees}.
Thus, cooperation via noisy link cannot increase the sum DoF of
 the MIMO interference channel whereas message sharing can increase  it.
 Nonetheless, it should be highlighted that both techniques may increase the sum capacity of the MIMO interference channel.
 \begin{rem}
Message sharing can increase  the sum DoF of the MIMO interference channel as well as its sum capacity.
\end{rem}

In \cite{vaze2012degree}, the  DoF region  of MIMO cognitive interference channel is obtained
when  CSIT is not available. Interestingly, it is shown that CSIT is not necessary for DoF-optimal performance
at certain antenna configurations, e.g., when $N_2\ge N_1 \ge L_2$

\section{Cognitive  Radio and 5G  Technologies }\label{sec:5G}

Wireless communication systems have undergone a revolution about once every decade.
Such a revolution leads to a completely new standard making a new generation of wireless networks.
Expected to  commercialized around 2020,  the 5th generation (5G) mobile networks
must support about 1000 times higher system capacity
than current 4G systems, as well as 10 times less latency, and about 100 times more devices.
To provide such a huge system capacity,
 three key approaches  have been suggested: network densification,
 adding a large quantity of new bandwidth, and increasing spectral efficiency.

Cognitive  radio is one of the  technologies that can, in conjunction with several other promising technologies,
address the spectrum scarcity problem. 
In this section, we  study the interplay between cognitive radio and emerging 5G technologies such as
massive MIMO \cite{larsson2014massive},  cloud radio access networks (cloud RAN) \cite{vaeziCRAN,quek2017cloud,wu2015cloud},
mmWave communication \cite{rappaport2013millimeter},  non-orthogonal multiple access (NOMA), full-duplex, etc.
The goal is to understand how each potential technology can be combined with the cognitive radio to increase the spectral  and energy efficiency of wireless systems.

\subsection{Interference Management in 1G-4G}\label{sec:ICmanagement}

Practical interference management approaches can be divided into two main categories:

\begin{itemize}
  \item\textbf{ignore interference:}
When interference is sufficiently weak then it is usually ignored by treating it as noise.
Such an approach deals with  \textit{signal levels}.
Treating interference as noise  is  proven to be optimal in achieving the sum capacity of
the interference channel at very weak interference regime.
   \item\textbf{avoid Interference:} To avoid interference, usually orthogonal multiple access methods such as
     frequency division multiple access (FDMA), time division multiple access (TDMA),
     code division multiple access (CDMA), and orthogonal frequency division multiple access (OFDMA)
     is employed. These approaches deal  with  \textit{signal space}. Strong interferers can also be avoided
   by decoding and canceling it. Using fractional frequency reuse (FFR) is another way
to  avoid interference in practical wireless networks.
FFR orthogonally allocates frequency  at the  cell-border regions in which intercell interference is usually  high.
   \end{itemize}

In 1G-4G wireless technologies, the above orthogonal strategies have been adopted for interference management.
While the underlay and interweave cognitive radio systems can operate with the above mentioned
multiple access  techniques, the overlay cognitive radio proposes an inherently different approach, as
it implies using the same frequency/time for cognitive and non-cognitive  users.
As such, the overlay cognitive radio requires non-orthogonal multiple access techniques,
as described in the following section.

\subsection{Cognitive  Radio and NOMA}\label{sec:CR-NOMA}

Wireless systems must provide service to multiple users concurrently.
 Multiple access is a technique that allows multiple users to share an allotted
spectrum (a channel) in an effective manner. Multiple access schemes are commonly
designed to share the channel  \textit{orthogonally}. For example,
multiple access schemes in 1G-4G cellular networks, i.e., TDMA, FDMA, CDMA, and OFDMA,
all are orthogonal multiple access (OMA) schemes. This is because in these schemes
access to the channel is orthogonalized in  time, frequency, or code domain. That is,
no two users share the same spectrum at  the same time or using the same code.
The rationale behind such orthogonal access methods is to avoid inter-user interference which, in turn, makes  signal detection  simpler.
However, due to this resource rationalization, OMA techniques  can support a limited number of users and have low spectral efficiency.
While exponentially increasing  number of  devices, mostly Internet of Things (IoT) devices, are being introduced to
wireless communication networks, there has been a flurry of research activity on new types of multiple access methods, random access methods, 
and waveform design  that can accommodate such massive number of  devices in 5G and beyond networks \cite{vaeziMA}.

Non-orthogonal multiple access (NOMA), in contrast to OMA,  is referred to  techniques that
allow to  scheduled multiple users over a single  resource.
NOMA can be realized in different domains, including in  the code and power domains \cite{nikopour2013sparse,vaeziMA,saito2013non}.
In the code domain, similar to CDMA, each user has its own code (spreading sequences) for sharing the
entire resource, but these codes are not orthogonal.
In the power domain, NOMA
exploits the channel gain differences between the
users for multiplexing via power allocation.

From the information-theoretic perspective, power domain NOMA is merely a new name for a well-established  theory.
The basic theory of NOMA has been around for several decades under the name of the \textit{broadcast
channel} (BC) and \textit{multiple access channel} (MAC) in  a single-cell setting, and \textit{interference channel} (IC) in
a multi-cell network \cite{Cover,ElGamal2011network,vaezi2019noma}.\footnote{Although optimal uplink and downlink transmit/receive strategies are unknown
for  multi-cell networks, in general, a combination of NOMA and OMA results in the largest achievable region  \cite{vaezi2019noma}.
} The new name, NOMA, is
coined to differentiate it from the  conventional  multiple access technique in 1G-4G wireless networks  such as  TDMA, FDMA, CDMA, and OFDMA.
It should be, however, mentioned that although the theory of downlink NOMA (BC) has been around since 1960's,
it has not been implemented mainly due to the complexity associated with successive interference cancellation (SIC) required at the
mobile handsets \cite{vaezi2019non}.  Today, with  the advance of processors
it is  possible to implement SIC at the user equipment.\footnote{It is worth mentioning that complex
user terminal capabilities, such as network assisted
interference cancelation and suppression (NAICS),  has been included in  3GPP LTE-A.} This has stimulated a large body of
research in academia and industry on NOMA for 5G.

Similar to IC and BC,  by definition,  \textit{overlay}  cognitive radio  networks imply non-orthogonal transmission,
as they let noncognitive and cognitive users use the same resource concurrently.
NOMA cognitive radio may, however, refer to the case where there are multiple noncognitive users or cognitive users.
   In any case, the theory of overlay cognitive radio  networks, discussed in Section~\ref{sec:main},
can be used to  design  effective transmit/receive strategies when power domain NOMA is in place.
Combination of these two technologies can bring further spectral efficiency in addition to other benefits of NOMA.
NOMA can be also
applied to  \textit{underlay} cognitive radio  networks to improve  the outage probability \cite{vaezi2019interplay}.


\subsection{Cognitive Radio and Other 5G Technologies}\label{sec:CR-NOMA}

During past several years, a number of other  technologies have been considered for inclusion in in 5G
in academia, industry, standardization bodies. This  includes, but is not limited to,
massive MIMO \cite{larsson2014massive},  cloud radio access networks (cloud RAN) \cite{vaeziCRAN,quek2017cloud,wu2015cloud},
 mmWave  \cite{rappaport2013millimeter} and  full-duplex \cite{zhang2015full} communication.
These technologies can be combined with the cognitive radio to increase the spectral  and energy efficiency of wireless systems.

While there has been significant attention
to combine these technologies with cognitive radio,
in most of these works cognitive radio operate either in the interweave
or underlay paradigms. Then, there is a big gap in combining these technologies
with cognitive radios that operate in overlay paradigm. We believe,
the introduction of NOMA to practical wireless networks will pave the road for
the implementation of overlay cognitive networks in future wireless networks.

\section{Future Research Directions}\label{sec:future}

Cognitive radio has rendered many traditional
problems in information and communication theory.
It has also uncovered new problems that need
research. It is a gold mine of research problems,
in particular, in terms of fundamental limits. In this section, we
list some of those problems. We also discuss the challenges in
bringing those results into practice.

\subsection{Open Fundamental Limits}\label{sec:open}
Here, we outline the open information-theoretic problems for  the
two-user, $\mathrm{K}$-user, and MIMO cognitive interference channels.

\begin{itemize}
          \item\textbf{Two-user cognitive interference channel:} In the discrete memoryless case,
          as discussed in Section~\ref{sec:2userDMC},
          the capacity of this channel is known when the cognitive user is more capable than the noncognitive user, i.e., $  I(X_1,X_2;Y_1) \leq I(X_1,X_2;Y_2)$.
         Otherwise,  when $  I(X_1,X_2;Y_1)> I(X_1,X_2;Y_2)$, the capacity region is open. For the Gaussian case,
         the capacity region is fully characterized for weak interference ($|a| \le 1$). Besides, the capacity region is
          known for part of the strong interference regime,  as discussed in Section~\ref{sec:2userG}. In the
          Z-interference case, the capacity is open only for $ \sqrt { 1 + P_1/(1+P_2)} < |a| < \sqrt{1 + P_1 }$, as can be seen from Table~\ref{table3}.
          We believe, in the above Gaussian cases, time-sharing can increase the achievable region similar to that of the interference channel described in
          \cite[Lemma 3]{vaezi2016simplified} and  \cite[Lemma 1]{vaezi2015limiting}, and the references therein. This technique has been applied to enlarge
          secrecy achievable region of the Z-interference channel in \cite[Lemma 3]{bustin2016secrecy},
          and it give a better region compared to  the TDM/FDM region, too. Time-sharing is expected to improve
          the gap between the inner and outer bounds and improve or theoretical knowledge about this channel.

         \item\textbf{$\mathrm{K}$-user cognitive interference  channel:} There are very few  capacity results for the $\mathrm{K}$-user channels, with $K>2$,
        including the  3-user channel.
         Characterizing  new capacity results and/or obtaining any insight into the optimal solution for these channels would be very valuable.
         Another possible direction is to find personably low gaps between the inner and outer bounds if it is not possible to  find the  capacity region. This
          can help gain insight on achievable schemes that are not far away from the capacity region.

         \item\textbf{MIMO cognitive interference channel:}
   In Section~\ref{sec:mimo}, we indicated that capacity region of the MIMO cognitive interference channel is open, even with full CSI.
   In contrast, the  DoF region  of this channel is known both with and without CSI, see \cite{huang2009degrees}
   and \cite{vaze2012degree}, respectively. A bridging step would be to work on GDoF of this channel.
\end{itemize}

\subsection{More Practical Setting}\label{sec:prac}
Apart from the DoF region with no CSI in the MIMO setting \cite{vaze2012degree}, the results
 we discussed in the previous subsection  depend on both the \textit{non-causal} knowledge of noncognitive users'
 message at the cognitive transmitter and  perfect CSI at both transmitters. When this is not the case, dirty paper coding,
  and interference mitigation techniques in general, may suffer
in terms of rate. Cooperative relaying \cite{han2009cooperative} is an interesting direction in relieving the non-causal message knowledge.
 In the case of imperfect CSI, other notions of capacity such as  \textit{ergodic capacity} or outage capacity
can be studied. Most of such results in the literature are either for underlay or interweave cognitive radio, but not overlay cognitive radio systems.
In addition, it is very important to understand the
tradeoff between the schemed that need learning the channel and interference that should be  mitigated and more practical
interference management techniques mentioned in  Section~\ref{sec:ICmanagement}.

\section{Conclusion}\label{sec:con}

This chapter has provided a high-level overview of
recent  information-theoretic results for overlay cognitive radio
networks, which we believe will become one of the  technologies driving
the evolution to future cellular systems.
The capacity region of the cognitive interference channel has been established
under several channel conditions, including the case where the cognitive receiver is
more capable than the noncognitive receiver. These results collectively demonstrate that
when cognitive users know the noncognitive user's
messages in a non-causal fashion,  achievable
rates largely increases for both users.
These results, however, depend on both the non-causal knowledge as well as
having perfect CSI at the transmitters.
It would be very interesting to understand the fundamental limits of this channel
under delayed CSI or no CSI, and use the insight in the design of practical wireless networks.
Specifically, the capacity region is open in most $K$-user
cognitive channels, including the MIMO case, even with
perfect CSI assumption.

We have also highlighted the large potential of combining
cognitive radio with 5G specific technologies such as
NOMA, massive MIMO, and cloud RAN  in terms of
spectral efficiency, energy efficiency, and low latency.
There are still several challenges ahead to realize the
full potential of the technology, both in theory and practice.
 This gives researchers
a rich  research area to work.


\end{document}